# Light induced magnetic order

T. Jauk[1], H. Hampel[1], J. Walowski[2], K. Komatsu[1], J. Kredl[2], E.I. Harris-Lee[3], J. K. Dewhurst[3], M. Münzenberg[2], S. Shallcross[4], S. Sharma[4,5], M. Schultze[1]

[1]Institute of Experimental Physics, Graz University of Technology; Graz, 8010, Austria.

[2]Institut für Physik, Ernst-Moritz-Arndt-Universität Greifswald; Greifswald, 17489, Germany

[3]Max Planck Institute of Microstructure Physics; Halle (Saale), 06120, Germany.

[4]Max Born Institute for Nonlinear Optics and Short Pulse Spectroscopy; Berlin, 12489, Germany.

[5]Institute for Theoretical Solid-State Physics, Free University of Berlin; Berlin, 14195, Germany.

**Heat and disorder are opponents of magnetism[1,2]. This fact, expressed in Curie's law established more than a century ago, holds even in the highly non-equilibrium interaction of ultra-intense laser pulses with magnetic matter[3–6]. In contradiction to this, here we demonstrate that optical excitation of a ferromagnet can abrogate the link between temperature and order and observe 100 femtosecond class laser pulses to drive a reduction in spin entropy, concomitant to an increase in spin polarization and magnetic moment persisting after relaxation back to local charge equilibrium. This both establishes disorder as an unexpected resource for magnetic control at ultrafast times and, by the provision of a purely electronic mechanism that does not involve reconfiguration of the crystal lattice, suggests a novel scheme for spin-based signal processing and information storage significantly faster than current methodology.**

Electronics is the behavior of numerous electrons controlled through electric fields and structures on the nanoscale, with the fundamental concept of an on/off switch determined by whether a current of electrons flows or does not. Nature, however, endows electrons with a second property, its quantized intrinsic angular momentum, the spin. This forms both the prototypical two-level system to serve as a bit in quantum computing and data storage, as well as presenting the possibility of utilizing transport of spin angular momentum without charge flow as means of signal processing, both features that would, if realized, radically transform the information technology landscape.

Magnetism in solids, however, is determined by the average over an ensemble of numerous electron spins that can point "up" or "down" along a quantization axis, placing control over their number randomness at the heart of attempts to utilize the spin degree of freedom. The thermodynamic aspect of heat as inimical to magnetic order was therefore identified early on by the pioneers of magnetism[1,2], a fact that finds its most modern incarnation in the inability of ultrafast laser pulses – that inevitably deposit heat into matter – to enhance, i.e. control, magnetic order at ultrafast times. Femtosecond laser pulses, that have otherwise shown profound control over the electronic degrees of freedom of solids up to sub-femtosecond response times[7–13], are found in numerous all-optical experiments[3–6,14–17] to simply reduce magnetic order as result of charge excitation. A promising approach to circumvent this obstacle is the indirect control of the magnetic state through cascaded light-charge (i) and charge-lattice (ii) coupling, transferring the ideas of polymorphism induced magnetic transitions into the time domain[18]. By virtue of selective excitation of coherent lattice vibrations by electron-, terahertz- or light-pulses switching and transient enhancement of magnetism have been demonstrated with the ultimate response time dictated by the speed of atomic motion inside the medium[19–21], considerably slower than the rapidity of electronic motion in response to an external stimulus.

Here, by employing ultrafast magnetic circular dichroism microscopy, we challenge the historical verdict that heat opposes magnetization and magnetization control directly via light-charge coupling is impossible. A femtosecond scale linearly polarized laser pulse is demonstrated to induce an anti-thermodynamic decrease in magnetic randomness, i.e., spin entropy, leading to substantial (30%) transient increase in the average magnetic moment with persistent enhancement after charge equilibrium is reestablished. In tandem with state-of-the-art theory, we show this increase to constitute the first observation of light-induced magnetic order (LIMO) in a solid.

**Space-time resolved magnetization sampling**

To inspect pathways to control spin order and create magnetic moment directly by light, we employ ultrafast laser driven threshold photoemission electron microscopy (PEEM) near the Fermi level that we equipped with spin-sensitivity employing two-photon magnetic circular dichroism (MCD) detection, as illustrated in Fig. 1a. This experimental scheme provides direct access to the spatial arrangement of magnetic domains with sub-200 nanometer spatial resolution and, synchronously, to the sub-picosecond temporal evolution of their energy-resolved electronic and spin dependent occupations. Magnesium oxide capped cobalt-iron-boron (CoFeB/MgO) on tantalum support serves as the sample for this study, a canonical alloy for spintronic applications in half of a magnetic tunnel junction (MTJ) configuration[22–25]. The structure is illuminated with circularly polarized, sub-200 femtosecond duration, ~10 µJ pulse energy laser pulses at a central wavelength of 515 nm that trigger spin-selective 2-photon photoemission (2PPE) from the ferromagnet due to the MCD effect. The photoelectrons propagate through the 2 nm thick MgO adlayer into a sub-micrometer spatially resolved electron flight time multichannel-plate imaging detector (see Methods section for sample & experimental details). Fig. 1a visualizes the magnetic domain network resulting from recording PEEM images for left- and right-handed helicity sequentially and computationally subtracting them. Picometer thickness control of the ferromagnetic wedge allows tuning the magnetic anisotropy from an out-of-plane configuration (perpendicular magnetic anisotropy, PMA) over a region with room-temperature stable skyrmions to in-plane magnetic domains (IMA)[25]. In the PMA region, the linear increase in the ferromagnetic (FM) thickness manifests itself in a gradual shrinking of the domain size, ranging from large extended patterns of millimeter size to nanometer scale magnetic domains. The domain size evolution exhibits an MCD contrast of around 2% (see Extended Data Fig. 1), highlighting the capability of studying buried interfaces micro-spectroscopically in photoemission geometry.

Optical excitation of the sample structure with controllable time delay $\Delta t$ with respect to the probing pulse is achieved by 150 fs duration, 65 degree angle of incidence, linearly polarized near-infrared laser pulses carried at a central wavelength of 1030 nm (1.2 eV photon energy), and pulse energy ranging between 0.15 and 0.45 µJ (0.65-2 mJ/cm² fluence). Contrary to a blurring of the domain pattern, an expectation fueled by Curie's law and the body of previous work on ultrafast laser induced demagnetization[26–30], Fig. 1b displays an increase in the MCD contrast on sub-picosecond timescale, shown here for photoelectrons detected in a +/- 100 meV energy window around -1 eV below the Fermi Energy $E_F$. This rapid and surprising increase in MCD contrast and associated magnetic moment recovers partially within a picosecond and settles to a long lasting percental increase. Strikingly, this long-lived enhancement is energetically stabilized as it outlasts the electronic relaxation occurring with a time constant of $\tau_{rel}^{el}$=440 fs extracted from the energy-resolved photoelectron yield measurement recorded simultaneously to the MCD detection. This first observation of a lasting increase in magnetic circular dichroism through illumination suggests the sample exhibits enhanced spin order and magnetic moment in response to the optical excitation – notably triggered by linearly polarized light that does not carry orbital angular momentum (OAM) or spin angular momentum (SAM) and in the absence of any signature of phonon excitation. The far-reaching implications of the opportunity to create magnetic order directly through light call for a close, multi-modal inspection.

**Band occupation and energy dependent magnetic moment**

The spin-resolved occupied density of states (DOS) of the saturated CoFeB/MgO interface immediately after photoexcitation is computed by time-dependent density functional theory (TD-DFT, for computational details see Methods section) and displayed in Fig. 2a[31,32]. The laser excites minority carriers into unoccupied states in the conduction band, reducing spin-polarization in the vicinity of $E_F$. This reduced moment and its partial recovery is observed as laser induced demagnetization in Fig. 2b panel I, and resembles the findings of experiments based on the magneto-optic Kerr effect, which is exclusively sensitive to the contribution of electrons in states at the Fermi surface to the magnetic moment[3–6,14–17]. A reduced spin polarization in this specific energy interval is, however, not necessarily indicative of a reduction of the overall magnetic moment, which results from the integrated majority/minority imbalance (i.e., the energy-dependent spin polarization $P_{Spin}(E)$, Fig. 2a) over all energy intervals (henceforth: global moment). The spin polarization features a byzantine energy dependence, in particular around -0.7 eV below $E_F$ our theory predicts the existence of a compensation point (CP) at which both spin states are uniformly occupied, and the spin polarization inflects. Laser induced depletion of minority carriers below the CP should consequently cause an increase in spin polarization, as confirmed by our experimental observation (Fig. 2b, panel II). This discord of experimental observations extracted in different energy intervals of the very same measurement underlines the need for energy-resolved experimental methods to explore time-domain magnetism, as the strongly modulated energy dependence of band occupations featured by our sample system appears as a general property of ferromagnetic systems[33–38].

We find sign and magnitude of the response of the spin system to optical excitation as suggested by the DOS for all energy intervals to excellently agree to the transient signals observed in the experiment (see Extended Data Fig. 4). We detect an MCD increase from 0.47% to 0.52% (relative change of 10%) in the energy interval below the CP featuring the largest enhancement and from 0.015% to 0.027% (relative change of 80%) globally (see Extended Data Fig. 5).

The change of MCD contrast as signature of the spin polarization in individual energy intervals, thus holds as a plausible consequence of light-induced carrier redistribution in the DOS so long as the charge transfer occurs quasi-instantaneous in comparison to the carrier relaxation through cascaded scattering channels. However, a transient increase in the global magnetic moment (Extended Data Fig. 5) cannot be accounted for within this picture. Optical excitations do not couple spin channels, while the spin-orbit interaction, which does, is universally seen in the ultrafast interaction of light and magnetic matter to result in a transient decrease in moment. Moreover, simultaneously with the dynamical evolution of the MCD contrast in individual energy intervals, we observe the CP (i.e., the energy at which we detect a balanced PE yield for both probe-light helicities) to feature a $\Delta E=120$ meV transient and a $\Delta E=40$ meV lasting shift (see Fig. 3a). TD-DFT calculation, in contrast, finds negligible CP shift for similar excited charge to that of the experiment. This disconsonance with the widely accepted framework within which ultrafast spin dynamics is understood calls, as we shall show, for a model that can account for the system transiting from partial disorder to a state of enhanced spin order.

The experiment further reports the build-up of the change in MCD contrast in response to optical excitation to evolve with the same time constant of 150 fs as the photoelectron (PE) yield in the respective energy interval. The number of photoelectrons detected in each energy interval is a direct and sensitive probe for the carrier occupation in this segment of the DOS (Fig 2b, panels I&II, yellow lines). Except for an autocorrelation artifact around time zero originating in the momentum streaking of probe-light photo-emitted electrons by the pump light, the delay time dependent PE yield reports on the time evolution of the carrier excitation and their return into ground state configuration. The excited carriers are found to relax with a time-constant of $\tau_{rel}^{el}$=440 fs and after 2 ps, the PE yield remains constant, indicative of the carrier relaxation into ground state configuration occurring slightly slower in our samples than the few-hundred femtosecond relaxation times reported for other ferromagnetic systems[33–35,39]. We note the absence of signatures for phonon excitation or lattice expansion in our experiment, despite photoelectron spectroscopy being a sensitive probe for lattice reconfigurations[40,41]

As shown in Fig. 2b, the MCD signal reversion ($\tau_{rel}^{MCD}$=750 fs) separates from the PE yield evolution and reaches a static value after approx. 3 ps - notably different from the initial yield in in all energy intervals. While carrier redistribution in the static DOS plausibly rationalizes the transient changes in spin polarization at early times (< 200 fs), the occurrence of (I) the residual change $\Delta MCD$ (see Fig. 2b) outlasting the carrier relaxation, (II) the difference in relaxation times of electronic- and spin-system and finally (III) the large transient CP shift are all irreconcilable with an interpretation of spin dynamics as driven by changing spin occupations within an approximately fixed electronic structure or via lattice reconfiguration and spin-lattice coupling. Au contraire, these facts together suggest significant and ultrafast light-induced renormalization of the electronic band structure. We scrutinize the interdependence of spin order and electronic structure by computing spin disorder in the Korringa-Kohn-Rostoker coherent potential approximation for substitutionally disordered alloys within the atomic sphere approximation (KKR-ASA-CPA). Fig. 3b displays the renormalization of the density of states as function of the concentration $x$ for an alloy model $Fe_x^{\uparrow} Fe_{1-x}^{\downarrow} Co$ representing spin disorder of the Fe sub-lattice in which a fraction *1-x* spins are unfavorably aligned. Disorder is seen to substantially alter the electronic spectrum, broadening both the exchange split $t_{2g}$ and $e_g$ majority bands, and implying an energy dependent spin polarization $P_{Spin}(E)$ strongly dependent on the state of magnetic order, Fig. 3c. The CP is particularly sensitive to spin disorder: when 25 % of the spins are unfavorably aligned ($x$=0.75), the $CP_{75\%}$ suffers a downward shift of ~ $\Delta E$=250 meV as compared to the saturated situation. A strongly renormalized electronic spectrum responding dynamically to the state of magnetic order conclusively rationalizes those experimental observations that appear starkly inconsistent with spin dynamics involving only changes in occupation numbers in a fully ordered system. Indeed, both the experimentally observed upward shift in energy of the CP by 120 meV and the transient increase in global moment by 30% are in quantitative agreement with this picture.

**Light induced order via electronic band renormalization**

The nexus between spin order and band renormalization allows us to put our conjecture of an increase in spin order in response to optical excitation to a final experimental test without recourse to MCD detection. This independent experimental verification rests on photoelectron spectroscopy directly probing the energy-dependent occupancies and is thus unswayed by possible complications due to intricate dependencies in magneto-optics[42].

Concomitant to the photo-induced change in occupation responsible for the transient change in moment, the renormalization of the DOS by a change in magnetic order implies a permanent modification of the photoelectron yield in different intervals of the DOS. The pump-probe delay dependency of the number of detected photoelectrons in the energy interval below the CP and at $E_F$ are displayed in Fig 2b (for other energy intervals see Extended Data Fig. 4&9). After approximately 2.5 picoseconds, quasi-static modifications to the PE yield can be observed with lifetime orders of magnitude longer than the ~100 femtosecond timescales reported for electronic relaxation and thermalization in comparable sample structures[43–45]. This state therefore represents an energetic ground-state situation where the carriers that were excited within the initial electronic structure relaxed into a modified DOS. Indeed, as predicted by our electronic structure calculation assuming an increase in spin order, we detect a lasting increase in PE yield above $E_F$ and a reduction in all other energy intervals probed in the experiment. Below the CP, we observe a change of $\Delta PE\ Yield$=-1.5% in reasonable agreement with the predicted change of available DOS of 6% (see Extended Data Fig. 9). Since the ferromagnetic alloy features a DOS with continuous energy spectrum, sub-picosecond scale relaxation into the electronic ground state associated with heating of the crystal lattice is expected. The lasting changes of the energy-resolved PE yields, otherwise inexplicable, therefore reveal a change in electronic structure that, in our sample system, commands the increase in spin polarization. The resulting increase in magnetic order can be quantified by evaluating the magnitude of spin polarization required in theory to effect a CP shift of $\Delta E$=40 meV as detected in the experiment (Extended Data Fig. 8). Here, we obtain a 5% enhanced concentration of spins aligned parallel to the quantization axis corresponding to a light-induced magnetic gain in average moment of $0.16\mu_B$ ($1.625\mu_B$ before and $1.785\mu_B$ after optical pumping).

Optically induced magnetic order implies unfavorably aligned Fe atoms switching their spin direction on a time scale of tens of femtoseconds. To probe the microscopic mechanism of this ultrafast change in spin configuration driven by light that neither triggers a reconfiguration of the crystal lattice nor couples to the charge carrier's angular momentum, we turn to a super-cell TD-DFT treatment (Fig. 4a, for computational details see Methods section). While interpretation of observations in ultrafast magnetization dynamics and spintronics often rests on theoretical treatment assuming an aligned spin concentration of $x$=1, full saturation in real systems at finite temperatures is an idealization. Here, spin-disorder is modeled assuming one Fe spin to be anti-aligned in a fixed-lattice super-cell extending over 8 atoms, equivalent to $x$=0.875, and we numerically model the charge flow triggered by the laser excitation as a majority spin-polarized current through this supercell (see Fig. 4b). In our experiment, this directional flow of charge originates from majority electrons that get excited into vacancies above $E_F$ of neighboring defect atoms. At the defect site, our modelling finds this local spin polarized current to rapidly excite additional carriers into vacancies above $E_F$ that are exclusively of minority character. The so established minority spin vacuum at the defect site presents an unstable electronic situation as minority spin holes are brought into energetic overlap with majority spin electrons, an imbalance that drives quick local redistribution of charge between majority and minority channels and so switching of the defect atoms magnetization causing a reduction of the global spin disorder (see Fig. 4a). Remarkably, the computational results in the supercell feature the upward shift of the CP after current induced switching as seen in the experiment (see Extended Data Fig. 10), and previously missing in the TD-DFT calculation that assumed full saturation of the moment. This *ab-initio* finding provides strong evidence that LIMO directly links light and magnetism in static crystal structures and should be universally observable in systems that exhibit spin disorder and occur regardless of the spin current's source.

**Concluding remarks**

This work advertises recognition of the intricate energy dependence of magnetic phenomena and spin disorder as a resource for controlling spin polarization of matter without the need for lattice reconfigurations. The enhancement of spin order prompted by the direct action of ultrafast light marks a substantial advance in electromagnetism and will expand to a new standard in spin-configuration dexterity in condensed phase systems. LIMO provides a general route for a significant speed-up of spintronic functionality and the exploration of novel spin-state based technologies at opto-electronic processing speeds as it obviates the necessity to modify the crystal lattice that render current protocols for magnetization control indirect and limit their maximum attainable speed. We anticipate far reaching consequences for spintronic applications and research into 2D magnets[46], spin-sensitive photochemistry[47], spin-selectivity of electron-transport in a chiral media[48] and spin-glass transitions[49].

**Methods**

**Sample preparation and characterization**

Samples with a variation of FM CoFeB thickness $t_{CoFeB}$ are grown as wedges with nominally picometer thickness control. The films are grown in a UHV cluster system under the same optimized conditions as for the fabrication of Ta/CoFeB/MgO MTJs[25] on thermally oxidized Si(100) substrates. A sample of Si(100)/ 500 nm SiO2/ 5-3.7 nm (wedge) Ta/ 1.20-1.56 nm (wedge) $Co_{40}Fe_{40}B_{20}$/ 2nm MgO is investigated. The Ta and CoFeB layers are prepared by magnetron sputter deposition using growth rates of 0.078 nm/s and 0.045 nm/s, respectively, in an Ar atmosphere at a pressure of $5 \cdot 10^{-3}$ mbar and a base pressure around $5 \cdot 10^{-10}$ mbar. To obtain a gradual thickness variation, the samples are mounted at a 50° angle to the sputter source normal. This geometric factor results in a 20% linear Ta thickness reduction, and most importantly a 20% linear CoFeB layer thickness growth along the substrate. The resulting thickness gradient for a CoFeB film with a center thickness of 1.2 nm is 0.028 nm/mm. The Ta film thickness of 5 nm in the center results in an opposite thickness gradient of -0.116 nm/mm across the substrate. The MgO layer is deposited in-situ by e-beam evaporation in the UHV cluster with a base pressure in the range of $10^{-9}$ mbar and a deposition rate of 0.02 nm/s. Ex-situ annealing is performed in a vacuum with base pressures below $10^{-7}$ mbar for 1 h at the final temperature. During the process of sample optimization, we tested several variations in the sample preparation and post-annealing. The investigated sample was annealed for 1 h at 200°C to initialize the solid-state-epitaxy process at the MgO/CoFeB interface as for the MTJs. Magneto-optical Kerr effect (MOKE) measurements provide local magnetization loops and allow for exact position identification of the PMA to IMA transition position on the wedged layer samples.

**Photoemission experiment**

The photoemission experiments were carried out with a NanoESCAII EF-PEEM (ScientaOmicron & Focus GmbH) under a base pressure of $5 \cdot 10^{-11}$ mbar. The energy resolution was set to 100 meV. As a light source, the second harmonic of a ytterbium-based and diode-pumped laser (30 µJ energy, 200 kHz repetition rate, 150 fs full-width-at-half-maximum (FWHM) pulse duration, 1030 nm wavelength) (Light Conversion, UAB, Pharos) was circularly polarized by a quarter-wave plate and was focused (f=350 mm) on the sample under an incidence angle of 25° creating a focal profile of 60 µm (FWHM). To visualize the magnetic domain network, PEEM images were subsequently recorded for left- and right-handed helicity and afterward computationally subtracted from each other. A Mach–Zehnder-type interferometer was used to introduce a delay between the NIR and the visible pulses. The beam profile of the NIR pump pulses on the sample had a size of 170 µm (FWHM). The spatial pump-probe overlap was continuously imaged by using the reflected beam from the sample and a CMOS camera.

**Data evaluation**

To retrieve the magnetic domain patterns, we record energy filtered PEEM images for left and right helicity light and calculate the MCD asymmetry as follows:

$$A_{MCD} = \frac{I_{\sigma^+} - I_{\sigma^-}}{I_{\sigma^+} + I_{\sigma^-}}. \qquad (2)$$

Possible illumination inhomogeneities are corrected before computing the contrast. The MCD contrast as a function of energy or delay time is determined by taking the average of the areas with opposite magnetization ($A_{+,-}$)

$$MCD(E_{kin}, \Delta t) = \frac{1}{2} \cdot (A_+ - A_-). \qquad (3)$$

Fig. 1b (lower panel) displays three snap shots of the stripe domain pattern representing the initial, the transient and the outlasting state of the MCD evolution upon photoexcitation. The MCD trace is extracted by applying a tailored mask to the individual MCD images, which reads out the mean asymmetry values as a function of the pump-probe delay.

To obtain more quantitative understanding of the ultrafast demagnetization/enhancement dynamics in the CoFeB/MgO, the temporal evolution of the normalized MCD ($MCD_{norm}$) signals were fitted with a function based on the three-temperature model (3TM)[17]

$$MCD_{norm}(\Delta t) = \left[1 - \left(A_1 \cdot \left(1 - e^{-\frac{\Delta t}{\tau_M}}\right) - A_2 \cdot \left(1 - e^{-\frac{\Delta t}{\tau_E}}\right)\right) \cdot H(\Delta t)\right] * G(\Delta t) \qquad (4)$$

Where $*G(\Delta t)$ presents the convolution product with the Gaussian laser pulse profile, $H(\Delta t)$ is the step function, and $A_{1,2}$ are constants. The two-time parameters $\tau_M$ and $\tau_E$ describe the dynamical process of demagnetization. $\tau_E$ characterizes the relaxation time of electron-phonon interaction where energy equilibrates the electron with the lattice. $\tau_M$ characterizes the demagnetization/enhancement time when energy is deposited into the spin system. Furthermore, to get insight into the temperature evolution of the electrons and the lattice, we applied the 3TM as described in Refs.[3,4] to the demagnetization curve of Fig. 2b (panel I). The electronic temperature increases transiently by about 230 K and thermalizes with the lattice to a temperature of 440 K, corresponding to an approximately 5% decrease in magnetization according to the temperature dependent study of Ref.[50]. Therefore, the observed increase in MCD contrast cannot be attributed to structural dynamics of the lattice caused by elevated temperatures.

**Static Characterization of the Ta/CoFeB/MgO system**
The linear FM thickness increase in a MgO-based MTJ layer system manifests in a magnetic phase transition from out-of-plane anisotropy (perpendicular anisotropy, PMA) to in-plane-anisotropy. For this study, we focused on the PMA region. Extended Data Fig. 1 displays how the picometer thickness control affects the hysteresis loops from magneto-optic Kerr effect (MOKE) measurements and the size of magnetic domains captured by MCD-PEEM. With increasing CoFeB thickness the loop shearing increases and, correspondingly, the hysteresis decreases, as depicted in Extended Data Fig. 1a. Concurrently, the magnetic domains gradually shrink from large extended magnetic domains millimeters in size to smaller sized labyrinth or maze domains (Extended Data Fig. 1b). From a domain wall boundary, the spatial resolution of the imaging technique can be extracted by fitting an error function as shown in Extended Data Fig. 1c which yields approximately 200 nm comparable to previous threshold MCD PEEM studies[51,52]. The corresponding fast Fourier transformation (FFT) images of the domain patterns reveal an exponential rise of the domain width periodicity when approaching the transition to in-plane anisotropy as shown in Extended Data Fig. 1d. Together with the linear increase of the FM wedge, a macroscopic sample position can be assigned to a layer thickness, making the system highly reproducible for experiments with complementary methods.

Extended Data Fig. 2 presents an overview of the spin-dependent electronic structure and its influence on the energy-resolved images of the magnetic domains. Split by a compensation point (CP), the spin-polarization (Extended Data Fig. 2a) deduced from DFT calculations is divided into a more majority and minority-dominated regime in the experimentally accessible energy landscape. Extended Data Fig. 2b outlines the measured 2PPE spectrum and the corresponding energy dependent MCD contrast of the investigated system. The sign change in the spin-polarization manifests itself in the energy-resolved images of the stripe domain pattern (Extended Data Fig. 2c).

## Time-Dependent Density Functional Theory

*Ab-initio* calculations are performed using a time-dependent extension of density functional theory (TD-DFT)[53]. Within TD-DFT dynamical evolution proceeds via the time dependent Kohn-Sham equation

$$i\frac{\partial \phi_{jk}(r,t)}{\partial t} = \left[\frac{1}{2}\left(-i\nabla - \frac{1}{c}A_{ext}(t)\right)^2 + v_s(r,t) + \frac{1}{2c}\sigma \cdot B_S(r,t) \right.$$
$$\left. + \frac{1}{4c^2}\sigma \cdot (\nabla v_s(\mathbf{r},t) \times -i\nabla) + \frac{i}{c}S(t)J\right]\phi_{jk}(\mathbf{r},t) \quad (1)$$

where $\phi_{jk}(\mathbf{r},t)$ are two-component Pauli spinor time-dependent Kohn-Sham orbitals with quasi-momentum $\mathbf{k}$ and state index j, $\mathbf{A}_{ext}(t)$ is the external laser field, written as a purely time-dependent vector potential, $\sigma$ are the Pauli matrices, $v_S(\mathbf{r},t) = v_{ext}(\mathbf{r}) + v_H(\mathbf{r},t) + v_{XC}(\mathbf{r},t)$ is the Kohn-Sham effective scalar potential, and $\mathbf{B}_S(\mathbf{r},t) = \mathbf{B}_{ext}(\mathbf{r},t) + \mathbf{B}_{XC}(\mathbf{r},t)$ is the Kohn-Sham effective magnetic field. The external scalar potential, $v_{ext}(\mathbf{r})$, includes the electron-nuclei interaction, while $\mathbf{B}_{ext}(\mathbf{r},t)$ is an external magnetic field which interacts with the electronic spins via the Zeeman effect. The Hartree potential, $v_H(\mathbf{r},t)$ denotes the classical electrostatic interaction. In this work, we employ the adiabatic local density approximation to quantify the exchange-correlation potentials, the scalar $v_{XC}(\mathbf{r},t)$, and the exchange-correlation magnetic field, $\mathbf{B}_{XC}(\mathbf{r},t)$. Finally, $\mathbf{S}(t)$ is the effective tensor potential and $\mathbf{J}$ is the tensor operator $\nabla \otimes \sigma$ that generates the spin current. This spin-current drives the light-induced magnetic ordering in CoFeB (see Fig. 4b).

## The Energy-Time Landscape

Extended Data Fig. 3 shows the dependence of the temporal evolution of the PE yield and the MCD contrast on the pump fluence and the breakdown of the spatial domain structure at high fluences. As the pump fluence increases, more carriers are depleted from the valence band, as observed in the PE traces -1.1 eV below $E_F$ (Extended Data Fig. 3a upper panel), the more electrons can be detected around $E_F$ (-0.1 eV, Extended Data Fig. 3b upper panel) after relaxation from higher occupied states in the conduction band. The corresponding changes of the MCD contrast for these two energy windows are depicted in the lower panels of Extended Data Fig. 3a&b revealing that the transient enhancement and quenching increase linearly with the pump fluence.

Exposing the system to a pump fluence around 2 mJ/cm$^2$ leads to a collapse of the domain pattern that crystallizes in a smaller domain network (see Extended Data Fig. 3c).

Extended Data Fig. 4 displays the pump-probe traces of the PE yield and MCD contrast for several energies compares them with the results of the TD-DFT calculation. The experiment reports a depletion of carriers for energies below and an increase in photoelectron yield above $E_F$ (Extended Data Fig. 4c) that is in good agreement with the early time changes in occupation predicted by TD-DFT (Extended Data Fig. 4b, upper panel). As a corollary of the charge redistribution driven primarily by minority carriers, the spin polarization shows a small increase below the compensation point (CP), the energy regime of predominant majority electrons, and a decrease above CP, the minority weighted energy window (see Extended Data Fig. 4b, lower panel). We observe transient trends of enhancements and losses in the MCD contrast (Extended Data Fig. 4d) in all energy windows, as predicted by TD-DFT.

Extended Data Fig. 5 displays the total change in MCD contrast energy integrated over the probed spectrum accessible by 2PPE. We detect a small negative contrast that changes sign upon interaction with the near-infrared pump pulse and gains a 400% transient and 80% outlasting increase in magnitude. The total photoelectron yield shows a maximum carrier depletion of approximately 5%. After 2 ps, the PE yield returns to its initial state, suggesting that the excited carriers have relaxed back into the valence band. However, this is not indicative of carrier relaxation into their original ground state, as the PE yields in the individual energy windows still report a lasting change after 3.5 ps.

**Disordered Local Moment Calculations**

Spin disorder is considered via an alloy model $Fe_x^\uparrow Fe_{1-x}^\downarrow Co$ employing the single site mean field coherent potential approximation, as implemented in Korringa–Kohn–Rostocker method in the atomic sphere approximation (KKR–ASA)[54–57]. We take the ideal B2 structure of CoFe with a lattice constant of 2.848 Å, employ a monkhorst-Pack k-mesh of 20 x 20 x 20, atomic sphere basis functions up to $l = 3$, and use a 16-point Gaussian mesh for the integration of the Green's function. Exchange-correlation effects are treated via the Perdew-Wang 96 LSDA. In the coherent potential approximation (CPA) treatment of the disordered local moment state, the magnetic moment on the

Co sub-lattice collapses to zero, a sign of the itinerant character of magnetism, and indicative of possible significant roles for on-site correlation and short-range order. We thus model the Co sub-lattice as ferromagnetic and restrict the presence of disorder to the Fe sublattice. In Extended Data Fig. 6 are shown the local moments as a function of the disorder on the Fe sub-lattice as well as (upper axis) of the average moment $m$, defined as $m = \frac{1}{2}(xm_{Fe\uparrow} + (1-x)m_{Fe\downarrow} + m_{Co})$, i.e. the total moment per atom in the unit cell (with the saturation moment then $m_0 = m(x=1)$).

The spin polarizations for various states of spin disorder on the Fe sub-lattice are shown in Extended Data Fig. 7a, revealing the profound sensitivity of the CP to the state of disorder (see Extended Data Fig. 7b). By comparing the CP position with the experimental extracted CP evolution, we assigned each point in delay time the degree of order and the global magnetic moment (see Extended Data Fig. 8). We calculated the changes in DOS and spin polarization as a function of delay time for several energy intervals and contrast them with the measured MCD contrast and the PE yield (see Extended Data Fig. 9). Below CP (Extended Data Fig. 9d, panel I), the spin polarization shows a transient and a lasting increase upon ordering, which is in excellent agreement with the temporal evolution of the MCD contrast (Extended Data Fig. 9c, panel I). Concurrently, the DOS change indicates a reduction of available states of approximately 6% in this energy interval after 3.5 ps, which reasonably matches the residual change in the photoemission yield of about 1.5%. Note, that the photoemission signal is not as pure as the differential MCD detection. The two-photon photoemission (2PPE) spectrum involves additional sources that do not contribute to the MCD contrast, such as secondary electrons or photoelectrons that originate from regions other than the metal/oxide interface. At the compensation point (Extended Data Fig. 9c&d, panel II), our model and the experiment reveal comparable trends for the spin polarization and the MCD contrast, including an early inversion in sign and almost vanishing magnitudes at long-time scales. As in panel I, the PE yield reports a lasting change of approximately -0.5%, consistent with the 5% decrease in the DOS. Above the compensation point (Extended Data Fig. 9c&d, panel III), the loss in MCD contrast after interaction with the pump-pulse is in line with the reduction in spin polarization upon band renormalization. Approaching the Fermi energy, carrier relaxation through cascaded scattering channels and, consequently, the elevated electronic temperature impedes the experiment observing the proposed band renormalization. Close to the Fermi level, spin ordering leads to an increase in available states in the minority channel at the expense of majority carriers, resulting in an increase in spin polarization and an overall reduction of available states (see Extended Data Fig. 9d, panel I). The experiment, however, reports an ultrafast demagnetization and a gain in photoemission yield because of carrier redistribution and relaxation (see Extended Data Fig. 9c, panel I).

**Effect of Spin Current on Magnetic Order**

In presence of magnetic impurities, a majority spin current strongly stimulates i) inter-site spin transfer (as in OISTR[58]), and ii) spin-orbit spin flips. These are ultrafast processes, capable of destroying the magnetic impurities on sub-500fs timescale. Upon optical excitation, many majority spins are transferred from the ferromagnetically aligned sites to the impurity. This majority spin current, acting as a local minority current on the defect atom, additionally creates many d-character minority holes below the Fermi level. This unstable electronic situation in the antiferromagnetic impurity drives many spin-orbit flips causing the switching of the defect atom magnetization and an increase of the total moment (see Extended Data Fig. 10a).

The spin-current-induced realignment of the impurity atom is shown in Extended Data Fig. 10b. Having switched alignment, the magnetic impurities experience no compulsion to flip back, explaining a long timescale increase in order. Extended Data Fig. 10c shows the DOS for disordered and aligned atomic moments, and Extended Data Fig. 10d shows the spin polarization in each case, highlighting the higher compensation point in the absence of the disorder. Computational details: We used the Elk code[32]: 272 k-points in the irreducible Brillouin zone; 20 empty states per atom; extra local orbitals above the Fermi level; one AFM aligned Fe atom per 8 atoms (x=0.875); SOC strength was increased by a factor of 2 because of the adjacent Ta atoms; spin current potential corresponding to electric field strength of $10^8$ V/m was applied matching the expected size for FeCoB[59].

**Data availability**

The data that support the findings of this study are included in the manuscript. All data, code, and materials used in the analysis are available from the corresponding authors upon reasonable request.

**Methods References:**

**Acknowledgements** Photoemission experiments were performed on the NAWI Graz core facility NanoPEEM. Financial support from Zukunftsfonds Steiermark (project number 9026) and BMBWF (HRSM2016) is gratefully acknowledged. K.K. thanks the Nakajima foundation for financial support. S. Sharma, J.K Dewhurst and E.I. Harris-Lee thank the DFG for funding through project-ID 328545488 & TRR227c Project A04. S. Sharma and S. Shallcross thank the Leibniz Professorin Program SAWP118/2021.

**Figures:**

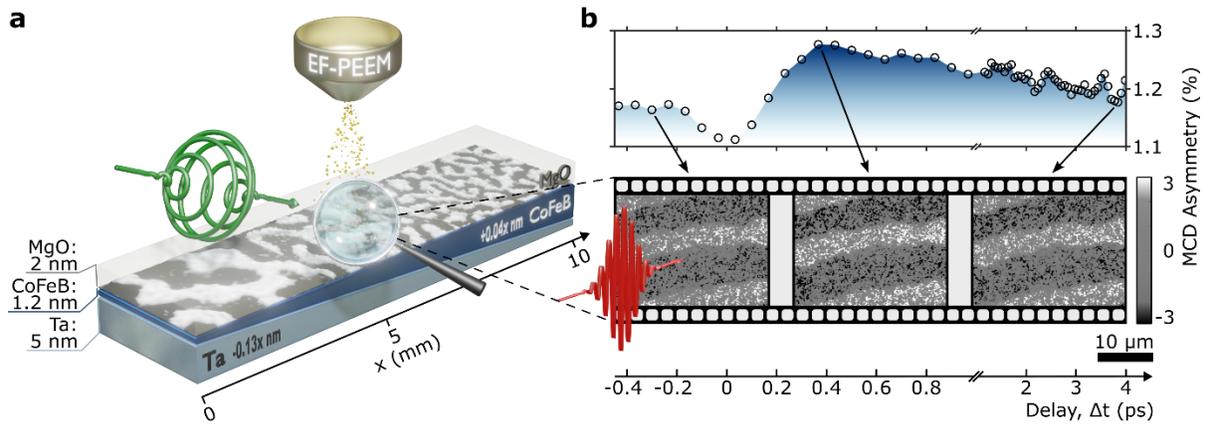

**Fig. 1| Light-induced enhancement of magnetic circular dichroism contrast. a,** Magnetic domain microscopy of a Ta/CoFeB/MgO trilayer, recorded by employing magnetic circular dichroism (MCD) in threshold two-photon photoemission (2PPE). **b,** Energy-resolved imaging captures an unexpected increase in MCD contrast (upper panel: line-out of MCD contrast) on a sub-picosecond timescale when optically excited with 150-femtosecond linear polarized laser pulses for an energy interval around -1 eV below the Fermi energy $E_F$ (lower panel: time dependent MCD microscopy recordings). This transient increase in contrast partially recovers, creating a long-term stable enhanced magnetic state that outlasts electronic carrier relaxation.

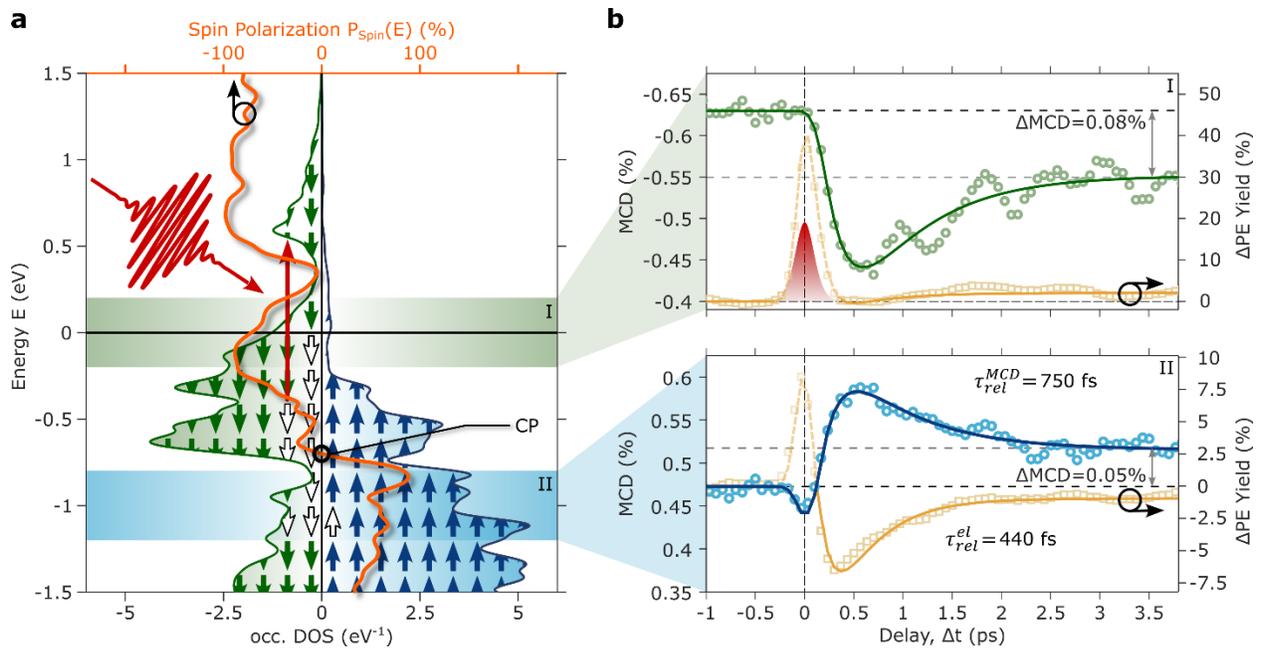

**Fig. 2| Band occupation redistribution and the time evolution of the magnetic moment. a,** The spin-dependent state occupation after photoexcitation computed by time-dependent density functional theory. The laser promotes minority carriers to unoccupied states above the Fermi level $E_F$, inducing an energy dependent response of the spin polarization. The depletion of minority carriers causes a reduced spin polarization around $E_F$ and an increase below the compensation point (CP). **b,** The measured evolution of the changes in photoelectron yield (yellow lines) and magnetic circular dichroism (green and blue line) for two energy intervals. While the minority-dominated energy window I (upper panel) close to $E_F$ exhibits a loss in magnetic moment, the moment in the majority dominated energy interval II (lower panel) increases after the interaction with the light pulse with ~ 150 fs duration symbolized be the red shaded gaussian bell curve).

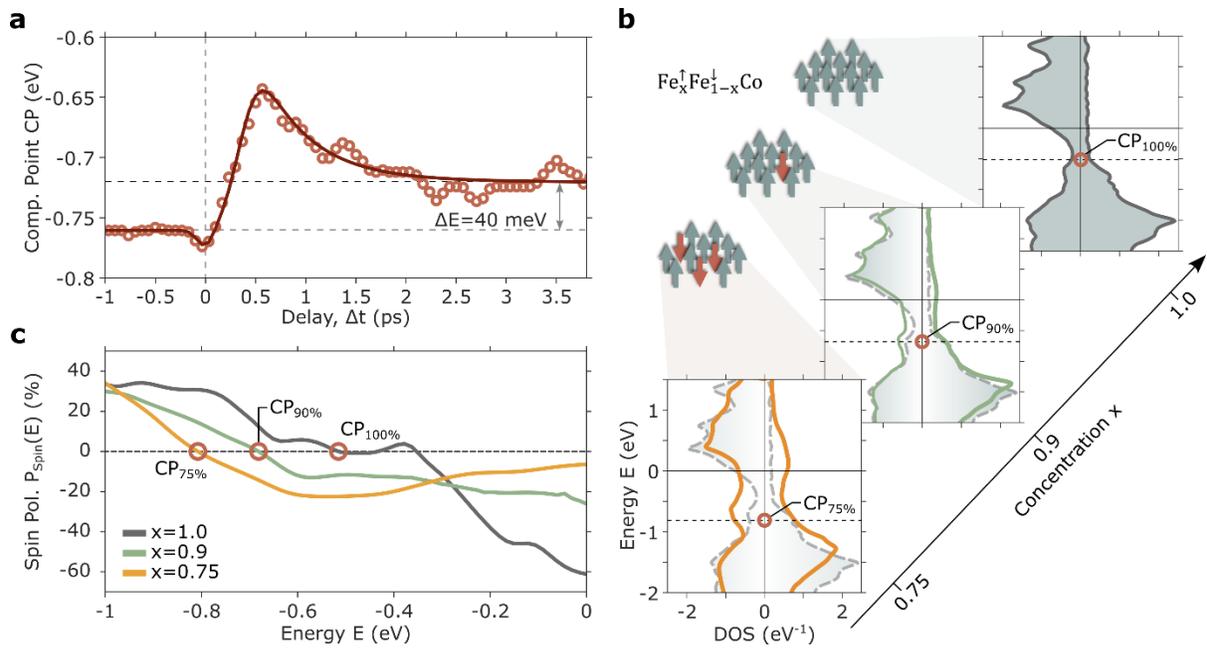

**Fig. 3| Band renormalization and spin disorder. a,** The experimentally observed time evolution of the compensation point CP featuring a lasting upshift in energy. **b,** The density of states as a function of the concentration $x$ for a spin alloy $Fe_x^\uparrow Fe_{1-x}^\downarrow Co$ representing spin disorder of the Fe sub-lattice in which a fraction of $1-x$ of spins are unfavorably aligned. **c,** The corresponding spin polarizations shown for fully ordered FeCo (gray line) and with concentrations of unfavorable Fe moments of 10% (green line) and 25% (yellow line). With increasing order, the compensation point (red circle) shifts towards the Fermi energy $E_F$.

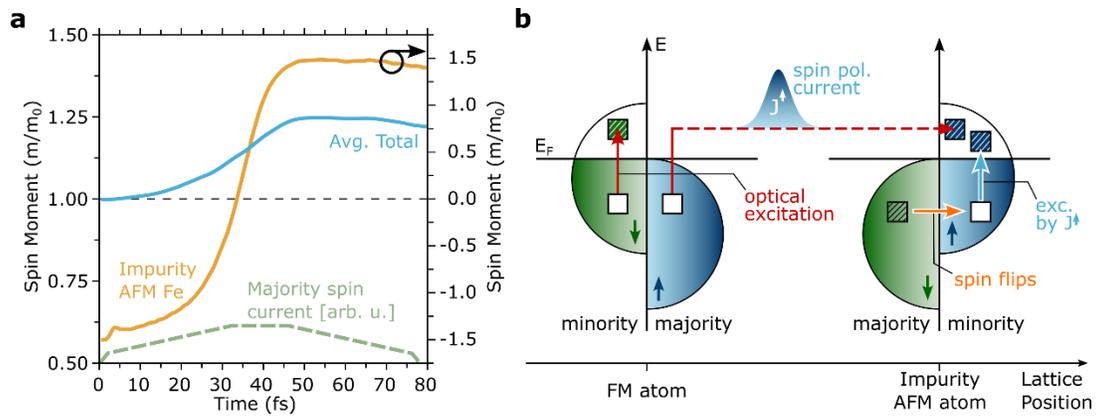

**Fig. 4| Light induced magnetic order. a,** Calculated time evolution of the average total moment (blue) and the moment of the impurity atom (orange) in an 8-atom supercell with one antiferromagnetically (AFM) aligned Fe atom in the presence of a spin current with majority polarization (green, dashed line). **b,** Laser excitation induces a majority spin-polarized current that acts as a local minority current at the impurity atom. This current excites further carriers, due to the absence of available states of majority type at the defect site exclusively creating minority vacancies. The occupation imbalance attracts majority carriers that transit into the vacancies undergoing spin-flips, finally toggling the defect sites magnetization direction.

**Extended Data Figures:**

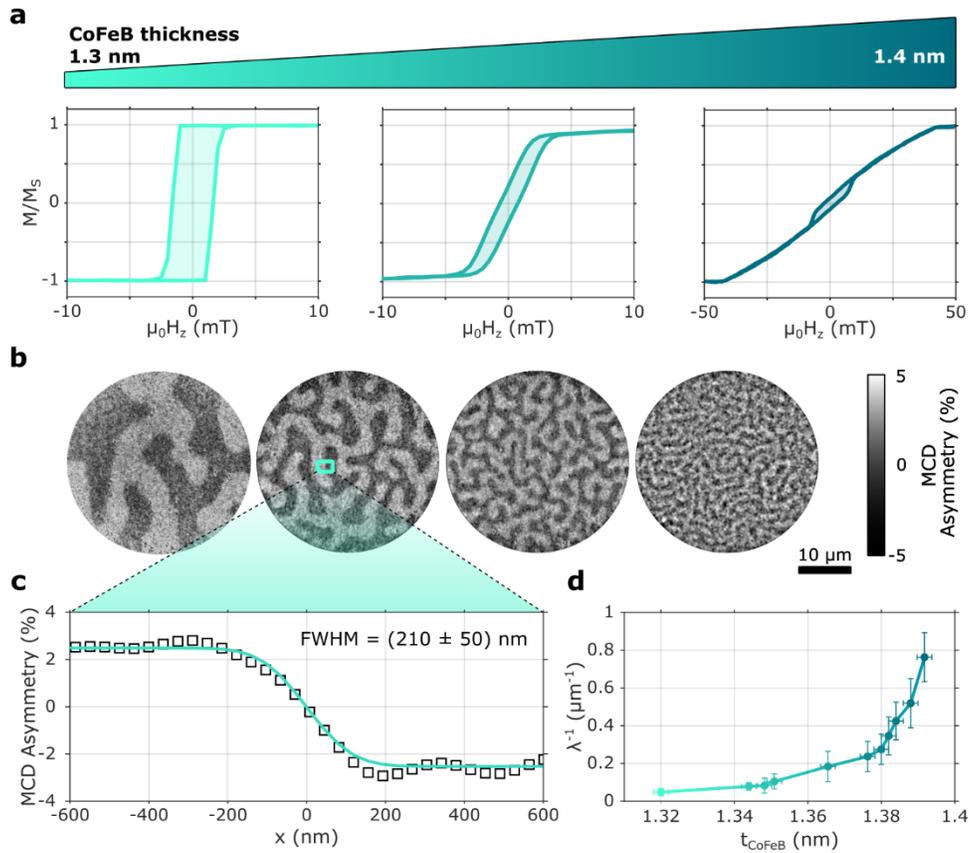

**Extended Data Fig. 1| Magnetic domain evolution of the perpendicular magnetic anisotropy region (PMA) and the imaging performance of the MCD photoemission technique. a,** MOKE hysteresis loops for different thicknesses of the CoFeB wedge showing the transition from the PMA region to the skyrmion phase. **b,** MCD-PEEM images of the domain size evolution ranging from large extended structures to labyrinth-like patterns. **c,** Line profile from a domain boundary resulting in a spatial resolution of around 200 nm. **d,** The periodicity of the domain width as a function of the thickness of the CoFeB layer.

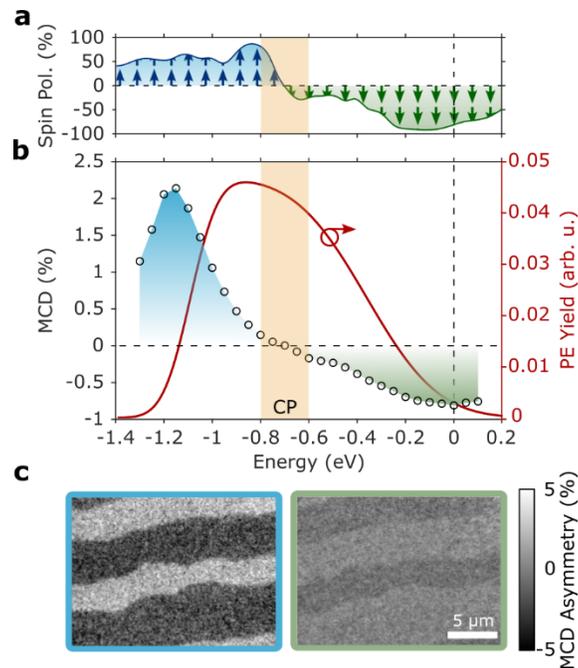

**Extended Data Fig. 2| Energy-resolved domain imaging near the Fermi level. a,** The spin-polarization deduced from DFT calculations of the CoFe/MgO system. **b,** Together with the two-photon photoemission spectrum (red solid line), the corresponding energy-dependent MCD contrast (open black circles) is plotted, also revealing a sign change of the magnetic moment in the probed energy landscape at the so-called compensation point (CP). **c,** The corresponding MCD images of the stripe domain pattern, integrated over the majority- (framed in blue) and minority-weighted (framed in green) energy windows.

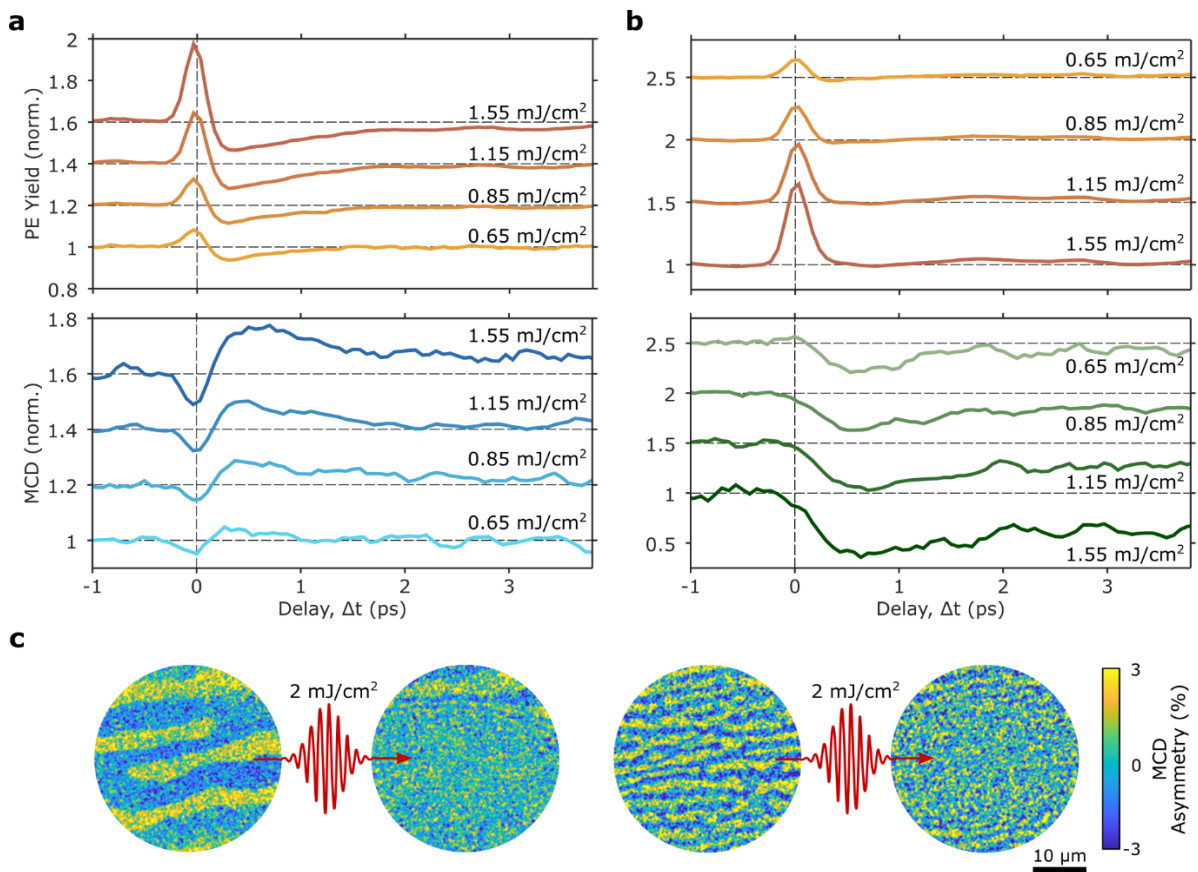

**Extended Data Fig. 3| Fluence-dependent occupation and moment dynamics. a,** The temporal evolutions of the normalized photoemission yield and the normalized MCD contrast for an energy window around -1.1 eV below the Fermi level for several pump fluences representing the majority weighted region below the compensation which shows the enhancing response. **b,** And the corresponding traces at -0.1 eV below the Fermi level representing the minority weighted window above the compensation point which shows ultrafast demagnetization upon photoexcitation. **c,** Increasing the fluence of the NIR-pump pulse leads to a breakdown of the domain pattern that crystalizes to a network of minuscule domains.

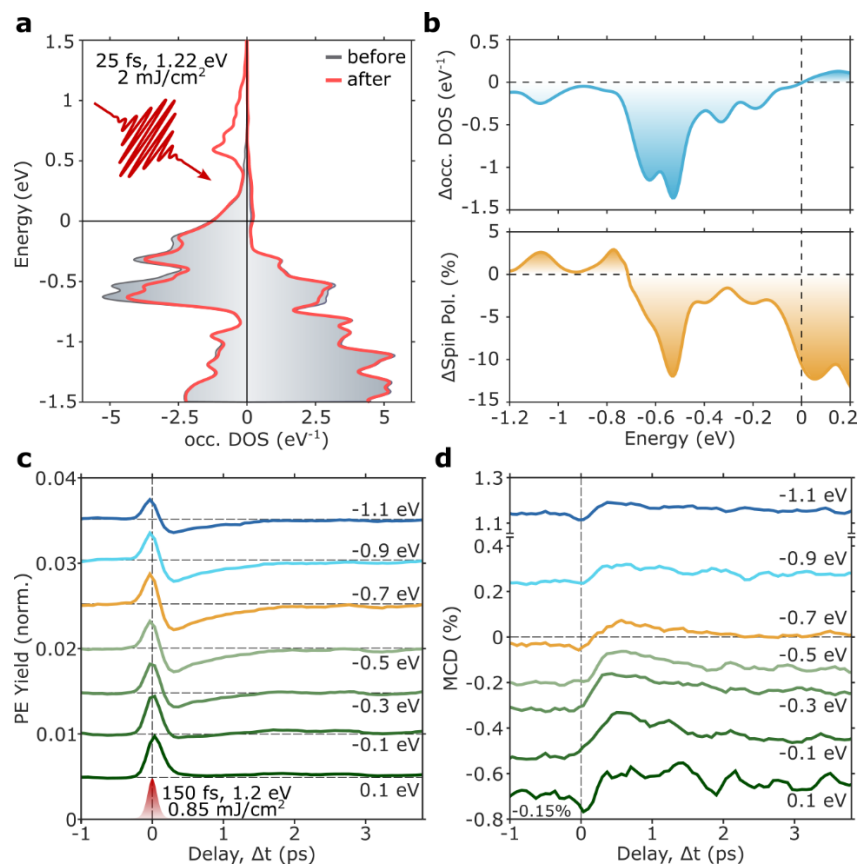

**Extended Data Fig. 4| Light induced band occupation redistribution and the spin system response. a,** The spin-dependent state occupation before (gray shaded) and after (red line) the interaction with a 25 fs near-infrared light pulse (1.22 eV, 2 mJ/cm²) computed by time-dependent density functional theory. **b,** The changes in occupation (upper panel) and spin polarization after photoexcitation. **c,** The normalized photoemission yield as a function of the pump-probe delay for various energy intervals of the valence band. **d,** The corresponding MCD contrasts extracted from the same energy intervals.

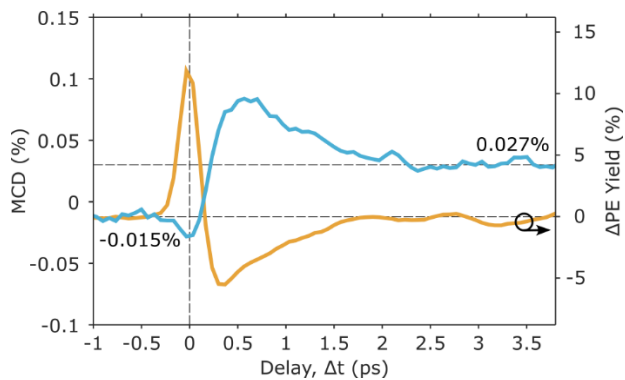

**Extended Data Fig. 5| The global evolution of the MCD contrast.** The MCD contrast (blue line) integrated over all probed energy intervals flips the sign after photoexcitation, transiently increases to 0.08% and outlasts the carrier relaxation (change in photoemission yield, yellow line) with an enhanced MCD contrast.

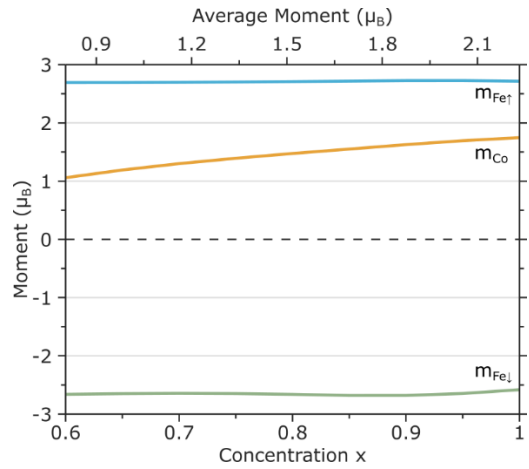

**Extended Data Fig. 6| The disordered local moment state in FeCo.** Shown are the local moments in a spin alloy $Fe_x^\uparrow Fe_{1-x}^\downarrow Co$ representing spin disorder of the Fe sub-lattice in which a fraction $1-x$ of spins are unfavorably aligned. The upper $x$-axis represents the average moment, $m = \frac{1}{2}(xm_{Fe\uparrow} + (1-x)m_{Fe\downarrow} + m_{Co})$.

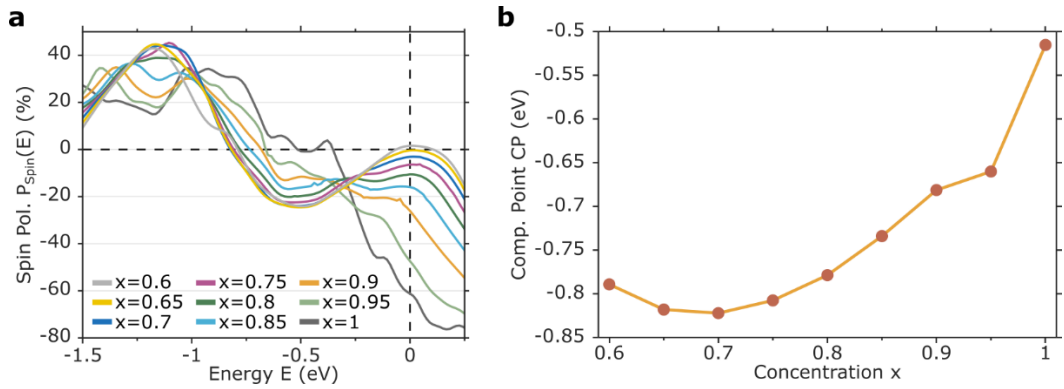

**Extended Data Fig. 7| Disorder evolution of the compensation point. a,** The spin polarization $P_{Spin}(E) = (D_\uparrow(E) - D_\downarrow(E))/(D_\uparrow(E) + D_\downarrow(E))$, presented as a percentage and as a function of energy for a range of spin disorder states of the Fe sub-lattice. **b,** The energy evolution of the compensation point as a function of disorder. Increasing $m/m_0$ towards saturation, i.e. decreasing spin disorder of the Fe sub-lattice, shifts the compensation point to higher energy.

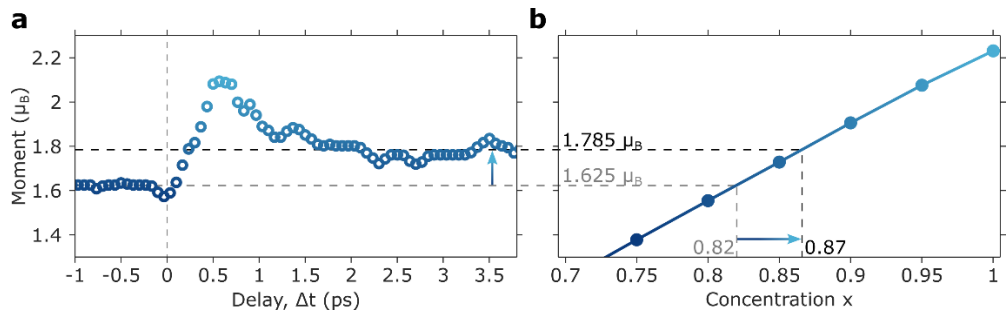

**Extended Data Fig. 8| The time evolution of the magnetic order. a,** The light-induced shift of the compensation point translates into an increase of the average magnetic moment by 0.16 $\mu_B$ (Bohr magneton), **b,** corresponding to a 5% enhancement in spin order.

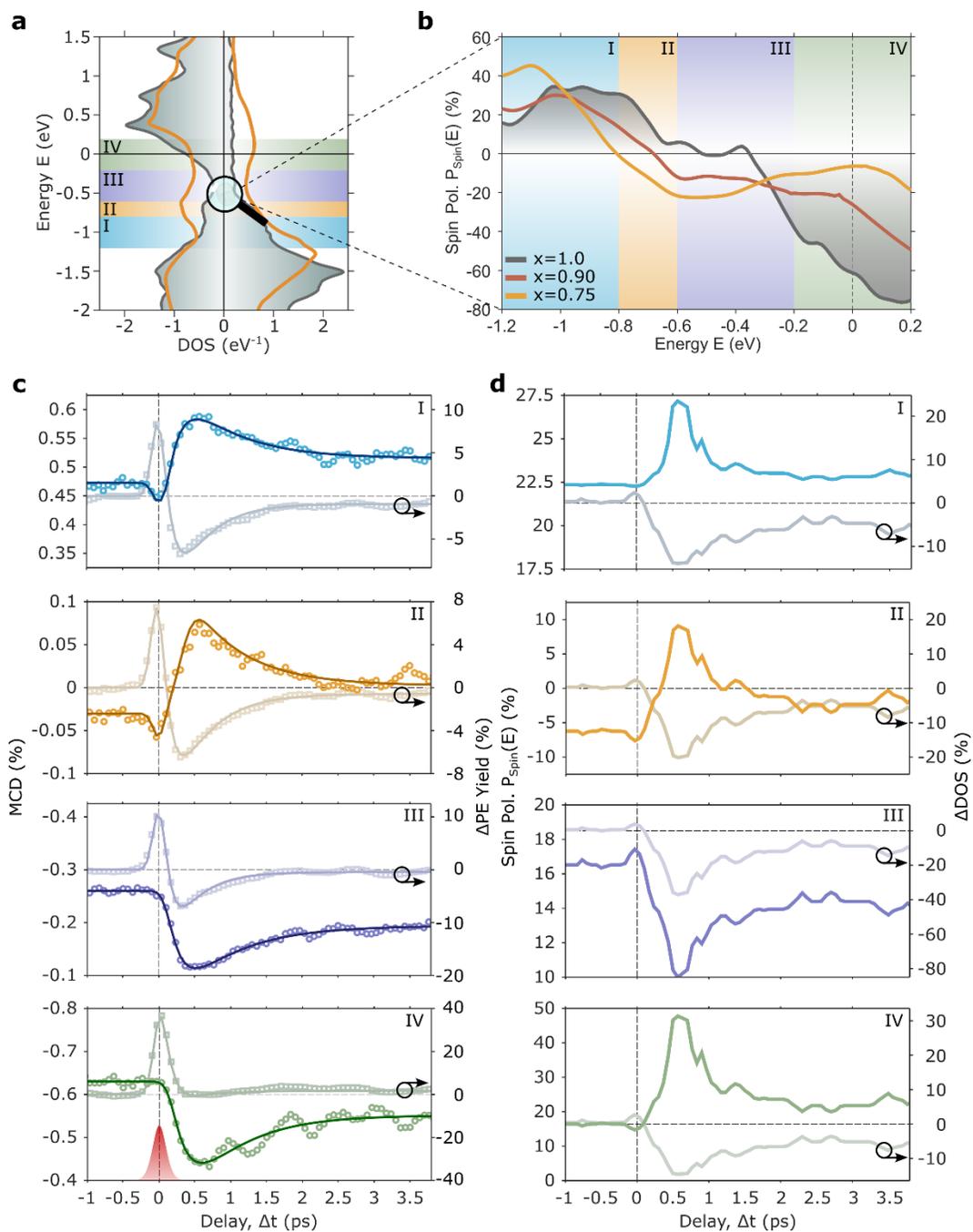

**Extended Data Fig. 9| Ultrafast band renormalization. a,** The density of states computed with the KKR-CPA-ASA methodology for an ordered spin alloy (gray shaded area) and a disordered configuration with a concentration of 25% spins unfavorably aligned (yellow line). **b,** The spin polarizations for the ordered and two disordered (x=10% and x=25%) configurations. **c,** The temporal evolutions of the MCD contrast and the changes in photoemission yield for the highlighted energy windows. **d,** The corresponding computed evolutions of the spin polarization and the changes in the density of states.

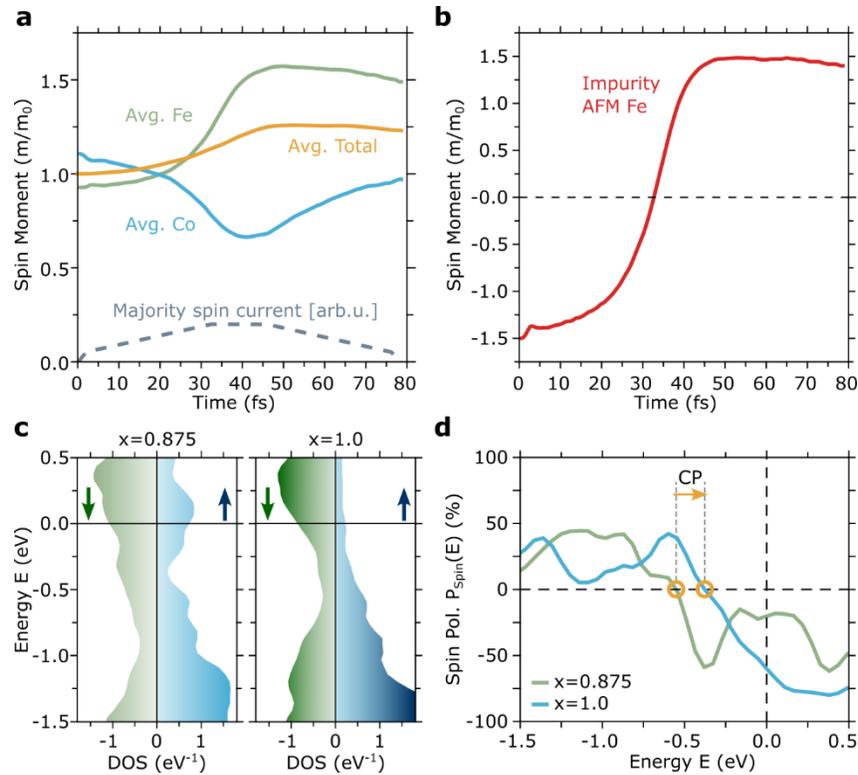

**Extended Data Fig. 10| Effect of spin current on magnetic order. a,** Calculated time evolution of the average Fe (green) Co (blue) and total (orange) moments in the presence of a spin current with majority polarization (grey, dashed line). $m_0$ is the initial average total moment. All of the atomic moments are initially ferromagnetically aligned, except for one Fe atom per 8 atoms, which is antiferromagnetically aligned. **b,** The time evolution for the impurity atoms. **c,** Density of states (DOS) for initial disorder (left, x=0.875) and for full magnetic ordering (right, x=1.0). **d,** The spin polarization for each of these DOS, showing the higher compensation point (CP) for the ordered state.